\documentclass{jgcc}


\usepackage{hyperref}
\usepackage{amssymb}
\usepackage{amsmath}
\usepackage[utf8]{inputenc}
\usepackage{float}
\usepackage{booktabs,caption,threeparttable}
\usepackage{array}

\theoremstyle{plain}

\begin{document}

\title{Non deterministic Pseudorandom Generator for Quantum	Key Distribution}

\author[A. Mishra]{Arun Mishra}	
\address{Department of Computer Science and Engineering, SoCE\&MS, Defence Institute of Advanced Technology, Girinagar, Pune, 411025, Maharashtra, India}	
\email{arunmishra@diat.ac.in}

\author[Kanaka R. P.]{Kanaka Raju Pandiri}	
\address{School of Quantum Technology, Defence Institute of Advanced Technology, Girinagar, Pune, 411025, Maharashtra, India}	
\email{raju@diat.ac.in}

\author[A.Pandit]{Anupama Arjun Pandit}	
\address{Department of Computer Science and Engineering, SoCE\&MS, Defence Institute of Advanced Technology, Girinagar, Pune, 411025, Maharashtra, India}	
\email{anupamapandit91@gmail.com}

\author[L. Sharma]{Lucy Sharma}	
\address{Department of Computer Science and Engineering, SoCE\&MS, Defence Institute of Advanced Technology, Girinagar, Pune, 411025, Maharashtra, India}	
\email{lucysharma95@gmail.com}

\begin{abstract}
  \noindent Quantum Key Distribution(QKD) thrives to achieve perfect secrecy of Onetime Pad (OTP) through quantum processes. One of the crucial components of QKD are Quantum Random Number Generators(QRNG) for generation of keys. Unfortunately, these QRNG does not immediately produce usable bits rather it produces raw bits with high entropy but low uniformity which can be hardly used by any cryptographic system. A lot of pre-processing is required before the random numbers generated by QRNG to be usable. This causes a bottle neck in random number generation rate as well as QKD system relying on it. To avoid this lacuna of post-processing methods employed as a central part of Quantum Random Number Generators alternative approaches that satisfy the entropy(non determinism) and quantum security is explored. Pseudorandom generators based on quantum secure primitives could be an alternative to the post-processing problem as PRNGs are way more faster than any random number generator employing physical randomness (quantum mechanical process in QRNG) as well as it can provide uniform bits required for cryptography application.\\
In this work we propose a pseudorandom generator based on post quantum primitives. The central theme of this random number generator is designing PRNG with non deterministic entropy generated through hard lattice problem - Learning with errors. We leverage the non determinism by Gaussian errors of LWE to construct non-deterministic PRNG satisfying the entropy requirement of QKD. Further, the paper concludes by evaluating the PRNG through Die-Harder Test. 
\end{abstract}

\maketitle

\section{Introduction}\label{S:one}

Quantum Key Distribution \cite{b18} strives for the strongest notion of security called as - Information Theoretic Security under classical authentic channel. The quantum mechanical process of QKD allows the detection of an adversary in the channel \cite{b18}. The protocol relies on quantum bits in form of two mutually independent basis. Either party chooses a $N$ random bits encoded in either of the basis on a random choice. This assignment of random basis is the key element of perfect secrecy in QKD systems and is derived from Quantum Random Number Generator(QRNG) \cite{b6}. By utilising the fundamental principles of random processes in quantum mechanics it makes frequent changes in the secret as required by One Time Pad \cite{b12} for perfect secrecy. The practical implementation of QRNGs have low overall eﬃciencies in terms of bit rate \cite{b11}. The fastest QRNGs till now are based on optical generators \cite{b5}. But generation rate is not the only problem with QRNG even the bits generated through them cannot be used directly for applications like cryptography \cite{b11}. The bits produced by them are raw and non uniform. The security of cryptography relies heavily on uniformity of the keys and thus the random bits generated by QRNG are not suitable for cryptography \cite{b10}. Currently, several post processing methods are applied to make the quantum random bits uniform and hence usable but these post processing methods render slow generation of bits even for high generation rate optical QRNGs \cite{b11}. This bottleneck can reduced if there is an eﬃcient post processing and quantum secure unit that can speed up with optical QRNGs and other faster QRNGS. \cite{b10}\\
Various post processing techniques are applied to the raw bits to distill quantum randomness such as extraction based on Topelitz hashing \cite{b9} which gave real time rate of 12gbps \cite{b10}. Generation rates upto 1 Gbps is achieved using extraction based on Homodyne detection \cite{b4}. Phase ﬂuctuation based random number generator produced upto 5.4 Gbps rate \cite{b22}. Another class of extractors that use pseudorandom generators are Trevisan extractors \cite{b19}. They use the underlying hard problem to generate uniform bits from quantum bits while having quantum security. Extractors based on the above pseudo random number generator (PRNG) \cite{b7} are faster than other paradigms of post processing, still there is a threat of attacks that can break the underlying hard problem. In addition to this, there is still a gap between generation rates of PRNGs and QRNGs at a higher degree. The fact that QKD protocol can detect the presence of adversary just by observing the quantum bits motivates us to design a random number generator that can be compliment the qualities of QKD protocol and further make its practical for cryptography.\\
This drives us to explore paradigms that oﬀer security at par with QKD. Recently, there has been a signiﬁcant progress in the study of cryptography primitives that are secure against adversaries equipped with quantum capabilities such as Lattice Based Cryptography \cite{b21}, Code based Cryptography \cite{b15}, Multivariate Cryptography \cite{b3}, and Elliptic curve isogeny based cryptography \cite{b8}. But not all of these schemes oﬀer practical use cases as its costs computational as well as memory resources and they are further limped by implementation complexities \cite{b2}. In the past few years, Lattice based cryptography has been taking a certain appeal due to provable worst case hardness guarantees \cite{b16}, its resistance from quantum attacks topped by ﬂexible implementations. These works have been mostly towards public key cryptography \cite{b18}, identity based encryption \cite{b19} and homomorphic encryption schemes \cite{b23}. Surprisingly Lattice based cryptography has very few exposure towards symmetric primitives such as Pseudorandom Functions(PRFs) and Pseudorandom Generators(PRNGs). Works by Banerjee et. al \cite{b1} on Pseudo random
functions aims to give generic constructions and direct constructions of PRNGs \cite{b1-1} from Lattices by utilising derandomizing techniques. Further, the derandomized form of the
PRF utilises Learning with Rounding (LWR) \cite{b1} problem which has proved to be directly reducible to Learning with Errors (LWE) \cite{b1} problem.\\
Our focus in this work is to derive a practical non deterministic pseudorandom generator based on provable hard problem called Learning with Errors hence we propose a non deterministic quantum secure pseudorandom generator for QKD . The non determinism and entropy of the approach is based on Gaussian error applied in Learning with Errors (LWE) Problem. The entropy of the PRNG is tested against Die Harder Test where it passes all the twelve test.
\subsection{Organisation}
The rest of the paper is organised as follows: In Section 2 we call the necessary preliminaries regards Pseudorandom Generators, Lattice based cryptography and its hard problems. In section 3 we introduce our approach of generating PRNGs from LWE. We demonstrate all the results of Security and Randomness tests. Finally, we preset the
application of the proposed PRNG in QKD.
\section{Preliminaries}
\subsection{Distinguishable advantage of a number generator}  
Given  a  number  generator  $g  :  \{0,1\}^m   \rightarrow  \{0,1\}^L$  with  $L  >  m$,  which  expands  an  $m$-bit
secret random seed into an $L$-bit sequence, we deﬁne as distinguisher in time $t$ for $g$ a probabilistic algorithm $A$ which, when input with an $L$-bit string, gives as result either $0$ or $1$ with time complexity limited by $t$. We deﬁne the advantage of $A$ for distinguishing $g$ from a perfect random generator as:
\begin{equation*}
    Adv_g(A)  = \|Pr_{x \in  \{0,1\}^m} (A(g(x) = 1) - Pr_{y\in  \{0,1\}^L} (A(y) = 1)) \|
\end{equation*}
The probabilities are considered over the values of a randomly chosen $x \in \{0,1\}^m$, a  randomly chosen $y \in \{0,1\}^L$, and the random choices of the algorithm A. We state that for distingushing the function $g$ in time $t$ as:\\

$Adv_g(t) = max_AAdv_g(A)$

\subsection{Pseudorandom Generators (PRNG)}
We consider a function $g$ to be a PRNG if $Adv_g$ is negligible for $t$ below a ﬁxed threshold. This allows to adjust the deﬁnition of a PRNG to the current accepted security levels.

\subsection{Lattice Based Cryptography}
Lattice intuitively can be seen in pattern making designs to crystallography and sphere packings. Informally, they can be thought of periodic arrangments of points in a Euclidean space. However after $1980s$ the computational aspects of these structures were investigated. In early $18^{th}$ century mathematician such as Gauss and Lagrange used lattice in number theory to give proofs of theorm. The theory was advanced mostly by Minkowski. Their application involved in solution of integer programming problems,
cryptanalysis, design of error correcting codes for multivariate systems and many more. During late $1990s$, for the ﬁrst time lattices were used to design cryptographic schemes as it is recognized as the source of computational hardness hence its application is designing secure cryptographic functions \cite{b14}.\\

\begin{figure}[h!]
	\centering
	\includegraphics[scale=0.95]{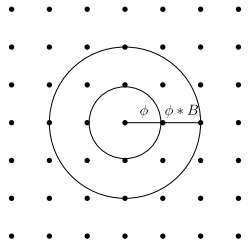}
	\caption{Discrete Points in a lattice}
	\label{fig:Lattice_Points}
\end{figure}

\begin{defi}  A lattice $\mathcal{L}$ is a $n$-dimensional discrete additive subgroup of $\mathbb{R}^n$. Additive subgroup implies that it is a group $< \mathbb{R}^n, + >$ and for any $x, y \in \mathcal{L}$ the following properties are satisﬁed \cite{b14}:\\

$x+y \in \mathcal{L}$    Presence of an Identity Element $0 \in \mathbb{R}^n$. Presence of an Inverse element $-x \in \mathcal{L}$ is present.\\

Discrete nature of lattices implies that within a distance period if there is a point, it should belong to the lattice. Formally, for every $x \in \mathcal{L}$ and $\phi > 0$, the point  $(x+\phi \ast B)\in \mathcal{L}$, where $B$ is the basis of the lattice. $Figure$ \ref{fig:Lattice_Points} represents a Lattice.
\end{defi}

\subsubsection{Hard Problems}
We introduce the most important problem in Lattice Based Cryptography: The Shortest Vector Problem (SVP) \cite{b20} which reduces to several other computationally hard problems. The practical implementation of Lattice based cryptography relies on the security of reduced problems from SVP - Learning with Errors (LWE) and Shortest Integer Solution (SIS) \cite{b13} problem as these problems gives worst case to average case reductions suitable for cryptography \cite{b20}.

\begin{itemize}
	\item Shortest Vector Problem: Given an arbitrary basis $B$ of some lattice $\mathcal{L} = \mathcal{L}(B)$, ﬁnd a shortest nonzero lattice vector, i.e., a $v \in (\mathcal{L})$ for which $|v| = \lambda(\mathcal{L})$ here
	$\lambda(\mathcal{L})$ is the length of the shortest vector in the lattice $\mathcal{L}$ \cite{b20}. Intuitively, SVP can be imagined as the ﬁrst point where the sphere touches as we increase it radius from the origin $O$. $Figure$ \ref{fig:SVP_Intitution} below demonstrates the intuition:
	\begin{figure}[h!]
		\centering
		\includegraphics[scale=0.7]{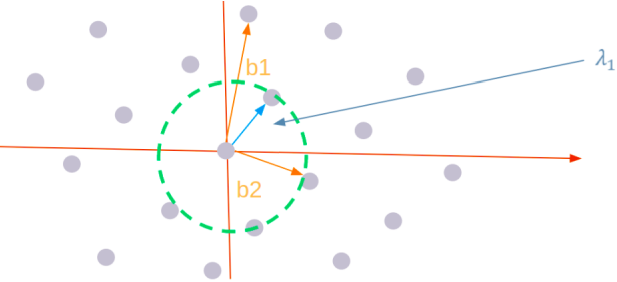}
		\caption{Shortest Vector Problem(SVP) Intitution}
		\label{fig:SVP_Intitution}
	\end{figure}	
	\item  Decision Approximate SVP: Given a basis $B$ of $n$-dimensional lattice $\mathcal{L} = (\mathcal{L}(B))$ where either $\lambda_1(\mathcal{L}) \le 1$ or $\lambda_1(\mathcal{L}(B)) \ge \gamma(n)$ \cite{b20}
	\item Shortest Integer Solution [24]:  Given $m$ uniformly random vectors $a_i \in Z^n$ forming columns of a matrix $A \in Z^m$  of norm $||z\le\beta||$ such that:
	
	\begin{equation}
		f_A(z) := Az = \sum_{i}
		a_iz_i = 0 \in Z^n_q
	\end{equation}
	
	Without the constraint on $||z||$ it is easy to ﬁnd a solution via Gaussian elimination \cite{b20}.
	
	\item  Learning  With  Errors(LWE):  We recall the learning with errors problems  which  claims  to  be  as  hard  as  worst  case  lattice  problems (SVP)  rendering  all cryptographic constructions based on it to be secure under the assumption that the worst case lattice problem is hard. We ﬁrst give an intuition of LWE problem with the example below then formally deﬁne LWE problem \cite{b17}.\\
	The LWE problem asks to recover a secret $s \in Z^n_q$ giving a sequence of ’approximate’
	random linear equations on $s$. Example the input could be:\\
	\phantom{...........}$5s_1 + 2s_2 + 7s_3 + 4s_4 \approx 9$\\
	\phantom{...........}$10s_1 + 6s_2 + 34s_3 + 65s_4 \approx 34$\\
	\phantom{...........}$55s_1 + 29s_2 + 72s_3 + 45s_4 \approx 7$\\
	\phantom{...........}$9s_1 + 10s_2 + 7s_3 + 13s_4 \approx 21$\\
	
	\phantom{.................................}\vdots\\
	
	\phantom{...........}$57s_1 + 42s_2 + 17s_3 + 4s_4 \approx 62$
	
	where each equation is correct upto some small error and the goal is to recover $s$. If the error is not there $s$ would be very easy because there are $n$ equations and $n$ variables, recovering $s$ could be done in polynomial time using Gaussian elimination. Introducing errors makes the problem diﬃcult. The Gaussian elimination algorithm takes linear combination of $n$ equations and hence it ampliﬁes the errors to unmanageable levels hence
	no information can be gathered from the results.
	\begin{defi}
	LWE:  For positive integer dimension $n$ and modulus $q \ge 2$, a probability
	distribution $\chi$ over $Z$ and a vector $s \in Z^n_q$, deﬁne LWE distribution to be $A_{s,\chi}$  to be the distribution over $Z^n_q  \times Z_q$  obtained by choosing a vector $a \leftarrow Z^n_q$ uniformly at random, an error term $e \leftarrow \chi$ and outputting $(a, b = <a,s> + e $ $  mod $ $ q )$ \cite{b17}.
    \end{defi}
	An algorithm solves LWE with modulus $q$ and error distribution $\chi$ if for any $s \in Z^n_q$  given
	an arbitary number of independent samples from $A_{s,\chi}$ it outputs $s$ with high probability.
	This ﬁnding of $s$ is the LWE-Search Problem \cite{b17}.\\
	For a certain modulii $q$ and Gaussian error distribution $\chi$ the decision LWE problem deﬁned as $LWR_{n,q,\chi}$ distinguishes with advantage non negligible in $n$ between any desired number of samples $m=poly(n)$ of independent samples $(a_i, b_i)  \leftarrow  A_{s,\chi}$  and the same number of samples chosen from uniform distribution $U(Z^n_q \times Z_q)$ is as hard as the LWE-
	Search Problem. For the mildest known requirements of $q$ ($q$ is a power of $2$) the above problem is hard \cite{b17}. LWE problem provides non-determinism in the output by incorporating random, independent errors.
	
	\item  Hardness of LWE:  The brute force method to solve LWE is through maximum likelihood algorithm. Assume for simplicity that $q$ is polynomial and that the error
	is in normal distribution. After assigning $O(n)$ values to equations the only assignment that satisﬁes the equation will be the correct one. This can be shown by a standard argument based only on Chernoﬀ’s bound and a union bound over all $s \in Z^q_n $. The algorithm uses only $O(n)$ samples and runs in time $2^{n\log n}$ \cite{b13}.\\
	There are several reasons to believe the LWE problem is hard. First, because the best known algorithms for LWE run in exponential time (and even quantum algorithms don’t
	seem to help). Most importantly, because LWE is known to be hard based on certain assumptions regarding the worst-case hardness of standard lattice problems such as GAP SVP (the decision version of the shortest vector problem) and SIVP (the shortest independent vectors problem). More precisely, when the modulus $q$ is exponential, hardness is based on the standard assumption that GAP SVP is hard to approximate to within polynomial factors \cite{b14}.
\end{itemize}
\section{Methodology}
This section describes the complete algorithm for the proposed non deterministic pseudorandom generator. The PRNG construction has two elements: Seed Hiding with LWE and LFSR sequence generation. The work ﬁrst applies LWE algorithm on the seed to prevent attacks aimed at retrieval of seeds. Secondly, LFSRs are instantiated using the secure seed to generate long sequences of random bits. This work is a direct construction of practical non deterministic PRNG from Lattice based primitive.

\subsection{LWE-Hiding Problem}
We now deﬁne the ‘LWE- Hiding problem’ which are modiﬁed version of LWE problem designed for the proposed Pseudo random generator.
 
\begin{defi}

Let $n \ge 1$ be the main security parameter and the prime modulus be $q$.

The decision variation of LWE-Hiding problem is as follows: For a given distribution over $s \in Z^n_q$, the $decision - LWE - Hiding_{n,q}$  problem is to distinguish between any desired number of independent samples $(a_i, b_i) \leftarrow U_s$ and the same number of samples drawn uniformly and independently from $Z^n_q \times Z_q$.
\end{defi}

\subsubsection{LWE-Hiding Problem Security}
We now show that for appropriate parameters decision-LWE-Hiding problem is as hard as decision-LWE.

\begin{itemize}
	\item \textbf{B-bounded distribution:} We say that a probability distribution $\chi$ over  over $R$
	is B-bounded if $Pr_{x\leftarrow \chi}[[x] > B] \le negl(n)$
\begin{thm}
     Let $\chi$ be any eﬃciently sampleable B - bounded distribution over
	$Z$ and let $q > B.n^{\omega(1)}$. Then for any distribution over the secret $s \in Z_q^n$ solving $decision-LWE-Hiding_{n,q,\chi}$  is as hard as solving $decision - LWE_{n,q,\chi}$.
\end{thm}
	
	\proof\hfill
    The central idea behind the reduction is simple: given pairs $(a_i, b_i) \in Z^n_q \times Z_q$
	distributed according to LWE distribution $A_{s,\chi}$. We prove the above theorem by contradiction. Let us assume that LWE-Hiding problem and LWE problem are distinguishable by a polynomial time adversary. We deﬁne two algorithms : 	$\mathbb{A}_{LWE_{Hiding}}(A, s)$ and $\mathbb{A}_{LWE_{Original}}(A, s)$ deﬁned as follows:\\
	
	\textbf{Algorithm 1:}  $\mathbb{A}_{LWE_{Hiding}}(A, s)$ \cite{b1} $r  \in \{0, 1\}^n $ Sample Error vector $e \leftarrow Z^m_q,  B: B = A \cdot s + e + r \cdot \frac{q}{2}(A, s, B)$
	
	\textbf{Algorithm 2:} $\mathbb{A}_{LWE_{Original}}(A, s)$ \cite{b1} $r  \in \{0, 1\}^n $ Sample Error vector $e \leftarrow Z^m_q,  B: B = A \cdot s + e (A, s, B)$
	
	Consider a polynomial time adversary $Adv$ querying $\mathbb{A}_{LWE_{Hiding}}(A, s)$ with $A, s$ sampled uniformly from $Z_q^{m \times n}$ and $s \in Z_q^n$. The algorithm returns $(A, s, B)$.
	
	\begin{equation}
		B = A \cdot s + e + r \cdot \frac{q}{2}
	\end{equation}

Since $r$ is a binary vector of length $r \cdot \frac{q}{2} + A \cdot s$ therefore $r \cdot \frac {q}{2}$ is very negligible hence equivalent to $B = A \cdot s + e$ which are samples when $Adv$ queries $\mathbb{A}_{LWE_{Original}}(A, s)$. This contradicts our assumption that both the samples are distinguishable. Hence it is proved that $LWE - Hiding$ distribution is indistinguishable from $LWE$ distribution.
\qed	
\end{itemize}

\subsection{Hiding Seed With Learning With Errors}

The initial phase of the proposed PRNG consists of hiding the seed using a hard function. Here the hard function is LWE problem of Lattice based Cryptography . The non-determinism required for applications like QKD is achieved through incorporating random and independent error in LWE problem. In addition to that LWE hiding is the central element required for security of the proposed PRNG generator. The construction utilizes the Lattice based hard problem called Learning with errors to attain security of the seed.\\
We ﬁrst describe the LWE based hiding function: LWE-Hiding(seed). Let $q, m, n$ be
integer parameters. To hide a seed $r \in \{0, 1\}^n$ sample a secret $s \in Z^n_q$. Choose a uniform $A \leftarrow Z^{m \times n}_q$ and sample an error $e \in Z^m_q$. Finally compute $b = A \cdot s + e + \frac{q}{2} \cdot r$ and output $b$, the seed hidden under LWE function. The pseudo code below describes the LWE-Hiding function.\\

\textbf{Algorithm 3:}  $LWE-Hiding(r)$ \cite{b1} Choose uniform $s \leftarrow Z^n_q$  Sample $A \leftarrow Z^{m \times n}_q$ Sample Error vector $e \leftarrow Z^m_q$, Hide seed $r: b = A \cdot s + e + r \cdot \frac{q}{2}$,
Hidden seed: $(b)$

\subsubsection{Non-determinism in LWE-Seed Hiding}

The Discrete Gaussian sampling is one of the most crucial components of Lattice Based Cryptography. It is the component that adds security by adding a random error term to
the matrix vector multiplication $A \times s$. The perturbation of the vector $A \cdot s$ contributes to the indistinguishability of the LWE samples from random samples. If the noise is not added LWE would give away secret information.\\

The errors are generated by taking Gaussian sampling over Lattice vectors. It samples
small vectors and performs addition to the $A \cdot s$ resulting in $A \cdot s + e$. The sampling is by
assigning probability of each lattice vector in the bell-curve. Hence for diﬀerent instances of time we get diﬀerent error resulting in non-deterministic result of $A \cdot s$.\\

In this work we leverage this non deterministic nature of LWE to design a random number generator that can inherit this property.

\subsubsection{Parameters and Implementation of LWE-Hiding}

Here, we describe the implementation details of the LWE-Hiding. For recommended
security the LWE-Hiding is computed under modulus $q = 8380417$ and $m$ and $n$ equal $4$. Therefore $A$ has a total of $16$ elements. Every element of $A$ (and $s$) is a polynomial in $Z^{m \times n}_q$$(Z^n_q$ for $s$).
\begin{figure}[h!]
	\centering
	\includegraphics[scale=0.58]{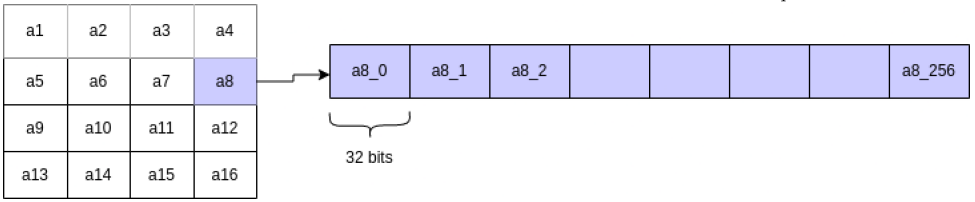}
	\caption{Coeﬃcient representation in matrix $A \in Z^{m \times n}_q$}
	\label{fig:Coefficient_Representation}
\end{figure}

These polynomials are represented in a construction of $256$ words, each word consisting of $32$ bits represents a coeﬃcient of the polynomial. Refer $Figure$ \ref{fig:Coefficient_Representation} for representation. Similar representation is designed for all the memory elements of LWE-hiding(). LWE-hiding returns the hidden seed $b$ which is a matrix of $m$ rows.
The implementation of LWE-Hiding consists of two major computations: Sampling the
matrix $A$ and multiplication of polynomials. To generate the samples of $A$ the proposed
work utilizes a symmetric scheme SHAKE-128. The short vectors $s$ is sampled using
rejection sampling and error $e$ is generated through Gaussian distribution between the
range $-1, 1$ to ensure that the perturbation by the error is short and it does not leak any secret.\\
For the main algebraic operation-multiplication of matrix $A$ whose elements are polynomials in $Z_q[X]/(x^{256} + 1)$ by the secret vector $s$ we consider Number Theoretic Transform (NTT) for low complexity upto $O(n \log n)$. NTT is just a version of FFT that works over the ﬁnite ﬁeld $Z_q$ rather than over the complex numbers. In our case, the school book method would take $4 \times 4 = 16$ polynomial multiplication. With NTT the multiplication reduces itself to point-wise multiplication which is very efficient for polynomials used in this work.\\
 The specific modulus $q = 8380417$ is chosen to provide a sufficient level of security against known classical and quantum attacks. The choice of $q = 8380417$ strikes a balance between security and efficiency. While a larger modulus might enhance security, it would also increase computational and storage requirements. The selected modulus is large enough to provide the desired security level while still allowing for efficient implementation and practical usage. The modulus $q = 8380417$ is deliberately chosen in the proposed scheme for efficient polynomial computations using NTT operations. However, it is important to note that $q$ is neither a power of 2 nor a large prime.\\
The selection process involves choosing an integer $k \geq 1$ and defining the modulus as $q = k \cdot 256 + 1$. This decision is based on the requirement of supporting the NTT operations and ensuring that $q$ is greater than or equal to the minimum working modulus necessary for the scheme.\\
It's important to note that the choice of modulus in cryptographic schemes is a result of extensive analysis, research, and consideration of various factors such as security, efficiency, and mathematical requirements. Different schemes may have different modulus choices based on their specific design goals and security considerations. The specific choice of $q = 8380417$ for the proposed scheme has been made based on these factors to provide an appropriate level of security while maintaining computational efficiency.

\subsection{LFSR Sequence Generation}
\begin{figure}[h!]
	\centering
	\includegraphics[scale=0.7]{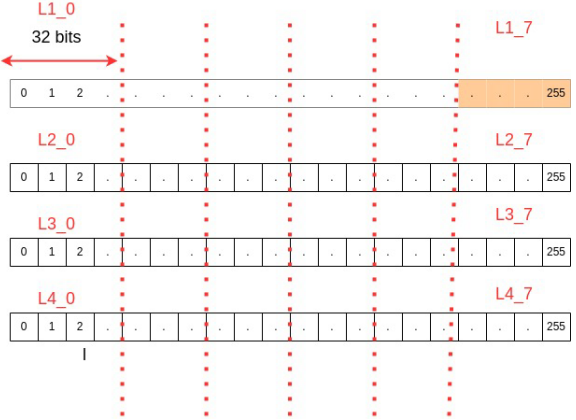}
	\caption{LFSR Construction}
	\label{fig:LFSR_Construction}
\end{figure}

Quantum Key Distribution aims to achieve perfect secrecy by utilising $n$ random bits for generating $n$ key bits. This indicates that the PRNG generator for the QKD applications should be able to generate millions of bits without exhausting itself. To achieve this the  second  step  of  this  work  employs  Linear  Feedback  Shift  registers  to  generate sequences indeﬁnitely. The input to the LFSR is the hidden seed - $b$ returned from LWE-hidding algorithm consisting of one polynomials. Each polynomial of $b$ consists of $256$ coeﬃcients and each coefficient is of $32$ bits. Therefore the total number of bits of $b$ is $256 \times 32 = 8192$. Running $4096$ bits around LFSR could have a huge impact on the performance of the PRNG as for one bit of shifting $4094$ bits would have to be shifted. This
will cause unnecessay delay in bit generation. Therefore we use only $1024$ bits of $b$ with each LFSR having $256$ bits, divided into $32$ bit coefficients. Each LFSR contains seven coefficients of the polynomial $b$. We take coefficients from the LSB and input it to our LFSR. We instantiate four LFSR each consiting of seven $32$ bit polynomial coefficents from $b$. $Figure$ \ref{fig:LFSR_Construction} shows the construction of $256$ bit LFSR.\\

The LFSR sequence generation consists of two algorithms:  Initialization  and  Feed-back / Output Generation. The initialization phase sets the LFSR by inculcating the bits from $b$ and arranges the eight coefficients into the void LFSR. This phase is monitored by a master LFSR which guarantees asynchronous filling of the LFSR. We choose a single Master LFSR $L4$ and all other $L1, L2, L3$ as slave LFSR. Master LFSR directs the shifting operation in feedback and output generation phase and also takes a central part in monitoring the filling operation of these LFSRs.

\subsubsection{Initialization Phase}
\begin{figure}[h!]
	\centering
	\includegraphics[scale=0.65]{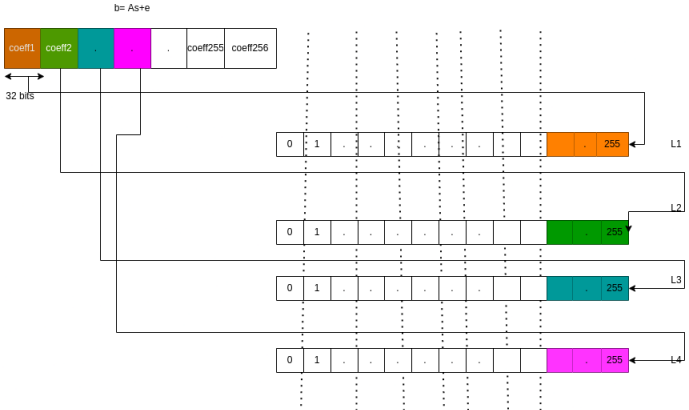}
	\caption{Initialization Phase of LFSRs: Filling First $32$ bits}
	\label{fig:Initialization_Phase_of_LFSRs}
\end{figure}

The initialization phase sets up four $256$ bit LFSRs $L1, L2,$ $L3, L4$. $L1 , L2, L3$
are  called  slave  LFSR  while  $L4$  is  the  master  as  it  monitors  the  filling  cycle  of  the
three  LFSRs.  The  filling  is  done  at  the  $32$-bits  each  that  is  LFSRs  are  ﬁlled  coefficient  wise.  At  first  the  $32$-bits  from  LSB  of  $L1,  L2,  L3$  and  $L4$  are  filled  from  the coefficients  of  $b$  $coeff_i$     $i  \in  \{1, 2, 3, 4\}$  respectively. Each  $32$  bit  coefficients  from  $b$
$coeff_1, coeff_2,$ $coeff_3, coeff_4$ is taken and filled into the first $32$ bits from LSB to LFSRS
$L1, L2, L3, L4$. $Figure$ \ref{fig:Initialization_Phase_of_LFSRs} represents the initialization Phase of LFSRs for filling first $32$
bits.\\

\begin{figure}[h!]
	\centering
	\includegraphics[scale=0.55]{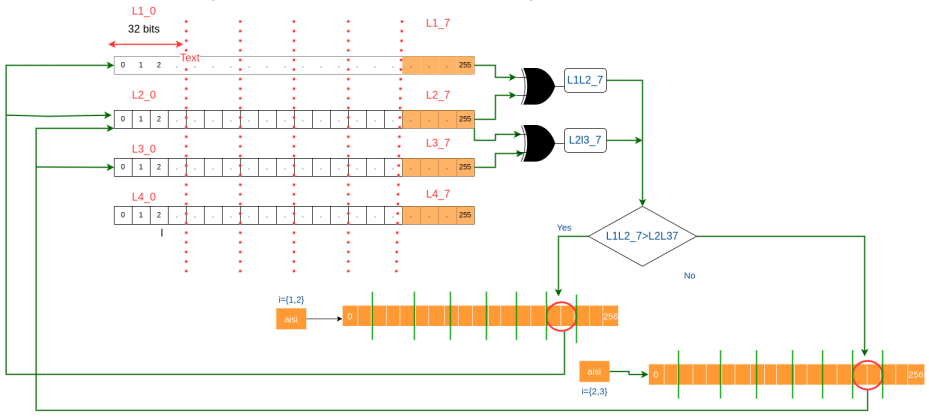}
	\caption{Initialization of LFSR: Filling Next $32$ bits}
	\label{fig:Initialization_of_LFSR_Filling}
\end{figure}

Next the cycle of filling the remaining $32$ bit slots of each LFSR starts for $L1, L2, L3, L4$ LFSRs. For each LFSR $Lji$ $j \in \{1, 2, 3, 4\}$ and $\in \{0 \le i \le 6\}$. The previously filled $32$ bit slots of LFSRs are $Lj7$ $ j \in \{1, 2, 3, 4\}$. In $Figure$ \ref{fig:Initialization_of_LFSR_Filling} the four $32$ bit polynomials
are highlighted: the coefficeint highlighted in orange gets filled in LFSR $1$ at the LSB bits, highlighted in orange, Similarly, the green coefficients are filled in LFSR $2$ at the
LSB, highlighted in green. The first $32$ bits of $L3$ and $L4$ are filled in similar fashion
highlighted with the colors green and pink respectively.\\

The next $32$ bits $Lj6$  $j \in \{1, 2, 3, 4\}$ is filled by first checking the highest among XOR
of $L17$  and $L27$ and XOR $L37$ and $L47$. The result of the LFSRs that are highest is filled
first. Suppose $L17$ and $L27$  is highest then $L1$ and $L2$ LFSRs would be given the chance
to fill ﬁrst. If the values are equal, all of the four LFSRs are given the chance of getting
filled at once irrespective of the results. $Figure$ \ref{fig:Initialization_of_LFSR_Filling} represents the initialization phase of LFSRs.\\

\begin{figure}[h!]
	\centering
	\includegraphics[scale=0.57]{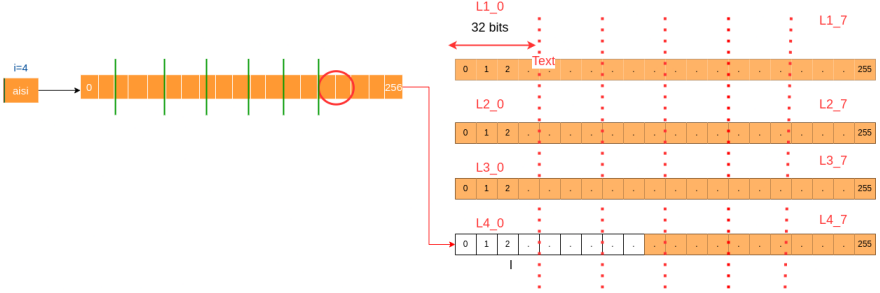}
	\caption{Initialization of LFSRs: Filling fourth LFSR ﬁlled}
	\label{fig:Initialization_of_LFSRs_Filling_fourth_LFSR}
\end{figure}

If  any  two  LFSRs  are  filled,  the  values  of  the  previous  coefficients  $Lji - 1 : j \in {1, 2, 3, 4}$ and $\in{0  \le  i  \le  6}$ are compared rather than the XORs. If three LFSRs are ﬁlled, the last LFSR is ﬁlled with the remaining bits of $b$. $Figure$ \ref{fig:Initialization_of_LFSRs_Filling_fourth_LFSR} represents the initialization Phase of fourth LFSRs.

\subsubsection{Feedback and Output Generation}
After the initialization phase, we have 256 bits LFSR cells ﬁlled with coefficient from
$b$ with each coefficient of $32$ bits. This phase takes $L4$ LFSR as the master and monitors the output and feedback 
generation of the three LFSRs. The LSB coefficient of $L4$ is checked and $L1, L2, L3$ LFSRs are shifted if and only if there is $'1'$ at $i^{th}$ bit position of the $32$ bit coefficient in $L4. L1, L2, L3$ are shifted by $i$ bits each. Let the shifted bits from $L1, L2, L3$ be $l1-o, l2-o, l3-o$. 
The feedback for $L1$ is $l1-o \oplus l2-o$, $L2$ is $l2-o \oplus l3-o$ and for $L3$ it is $l3-o \oplus (32$ bit of $L4)$. This cycle continues until all the ones of current coefficient in $L4$ is exhausted.\\

\begin{figure}[h!]
	\centering
	\includegraphics[scale=0.77]{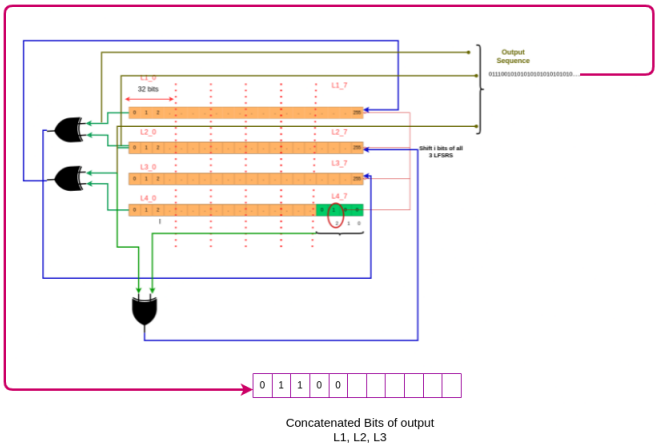}
	\caption{ Feedback and Output Generation by L1, L2 and L3}
	\label{fig:Feedback_and_Output_Generation}
\end{figure}

The shifted bits are the output generated by the PRNG. Each LFSR at one output cycle generates $32$ bits at once. The shifted bits generated by $L1, L2$ and $L3$ are taken
to the concatenation module which merges all the bits for the generation of ﬁnal output bits. In $Figure$ \ref{fig:Initialization_of_LFSRs_Filling_fourth_LFSR} the pink lines indicate the shifted bits taken to the concatenation module. Note that until now the output bits of LFSR $L4$ is not generated. $Figure$ \ref{fig:Feedback_and_Output_Generation} shows the feedback generation process for $L1, L2, L3$ mastered by $L4$ and the output sequence generation.\\

\begin{figure}[h!]
	\centering
	\includegraphics[scale=0.82]{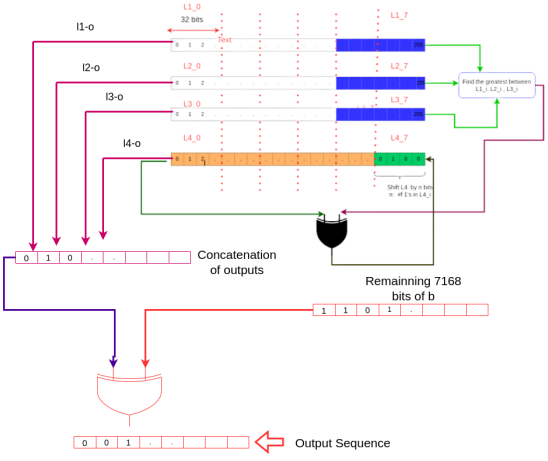}
	\caption{LFSR L4 Feedback and Output Generation}
	\label{fig:LFSR_L4_Feedback_and_Output_Generation}
\end{figure}

At the completion of each cycle of shifting by a coefficient in $L4$, the output and feedback for $L4$ is generated before moving to the next coeﬃcient. $L4$ is shifted according
to the number of $1's$ in the previous $32$ bits of the coefficients. The feedback for $L4$ is calculated by taking the highest value among the current coefficients and xoring with the shifted bits of $L4$. $Figure$ \ref{fig:LFSR_L4_Feedback_and_Output_Generation} shows the feedback generation for $L4$ LFSR. These shifted bits are concatenated with the output generated by LFSR $L1, L2$ and $L4$ indicated by pink lines in $Figure$ \ref{fig:LFSR_L4_Feedback_and_Output_Generation}.\\

All the four LFSR $L1, L2, L3, L4$ bits are concatenated. The number of bits generated by all the LFSRs $L1, L2, L3, L4$ is given by the following formula:\\ 

Total Number of bits generated by $L1, L2, L3, L4$ per cycle 
\begin{equation}
	= 3 \times \sum_{i = 1} i = 32p \times i + p
\end{equation}

where $p$ is the position of bit $1$ in $L4$ and $i$ is the index of coefficients $\in$ $\{1 \dots 32\}$ . Since
the coefficient bits of $L4$ governs the shift and hence the shifting bits of $L1, L2$ and $L3$ $3 \times \sum_{i = 1} i = 32p$ indicates this quantity. The master LFSR $L4$ Shifts according to the number of $1's$ in its $32$ bit coefficient per cycle indicated by $p$.
These bits are then XOR-ed with the remaining $7168$ bits of vector $b$ to generate the
ﬁnal output sequence. In $Figure$ \ref{fig:LFSR_L4_Feedback_and_Output_Generation} XOR indicated in red performs the operation with
remaining $7168$ bits of $b$ to give the ﬁnal output.

\subsection{Security}
The security of our scheme relies completely on the hardness of Learning With Errors in the Random Oracle Model (ROM). The hardness of standard LWE problems asks to distinguish between $(b = A \cdot s + e)$ from $(A \cdot s)$ sampled from uniform distribution. The proposed schemes utilizes the LWE Gaussian error addition to $A \cdot s$ along with the original
seed. The security of our scheme also relies on Shortest Integer Solution (SIS) problem. SIS seeks to ﬁnd short vectors such that A.s=0 where A and s are uniformly random. In
ROM ﬁnding such short $s$ for $A \cdot s + e + r \cdot \frac{q}{2}$ is a hard problem. In Quantum Random
Oracle Model where adversary can query in superposition the reductions for Module LWE suggests its security in QROM. The proposed LWE based hiding function relies
on module LWE with parameter $q$ such that its security is guaranteed in QROM. Hence by hiding the seed with LWE function, the adversary has to ﬁnd short vectors from $A \cdot s + e + r \cdot \frac{q}{2}$ or has to distinguish it eﬃciently. Both of the problems coincide to SIS
and LWE problem respectively. Hence the security of the seed can be guaranteed.
\subsubsection*{Key Space Analysis}
In the proposed PRNG, the input is obtained from Shake256. The Shake256 algorithm has a key space of $2^{256}$, which implies that there are $2^{256}$ possible keys that can be used as inputs.\\
The key space refers to the range of unique and distinct keys that can be utilized by the PRNG. With a key space of $2^{256}$, the LWE-based PRNG benefits from an incredibly large number of potential keys, ensuring a wide variety of choices for selecting the initial key or seed.\\
The size of the key space is of paramount importance for the security and unpredictability of the generated pseudorandom sequence. A larger key space makes it extremely difficult for an attacker to exhaustively search or brute-force the key space, ensuring the robustness of the generator's security.\\
Overall, the analysis reveals that the proposed LWE-based PRNG possesses a key space of $2^{256}$. This indicates a vast range of potential keys, providing a high level of security and resilience against attacks aimed at compromising the generator's randomness and predictability.

\subsection{Randomness Tests}
\begin{figure}[h!]
	\centering
	\includegraphics[scale=0.8]{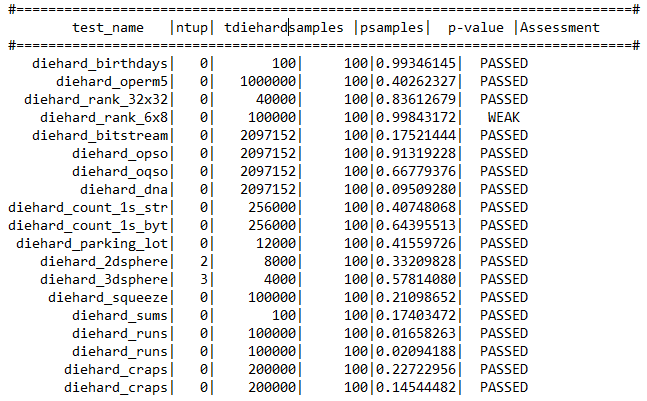}
	\caption{Die hard Test Results}
	\label{fig:Diehard_Test_Results}
\end{figure}

The experiments were conducted in the Secure Systems Lab, which is part of the Computer Science and Engineering Department, SoCE\&MS at DIAT, Pune. The CDAC-PARAM Shavak machine was utilized for performing the tests.  
The developed pseudo random generator was subjected to the Dieharder test suite to evaluate the generated entropy. To facilitate the analysis, the output bits, with a total size of 1.1 GB, were stored in a file. These buffered bits were then used as input for the Dieharder test suite. The test suite consisted of 12 tests, and a sample size of 10,000 was used for the analysis. The table in $Figure$ \ref{fig:Diehard_Test_Results} below shows the test that have been passed.

\begin{figure}[h!]
	\centering
	\includegraphics[scale=0.80]{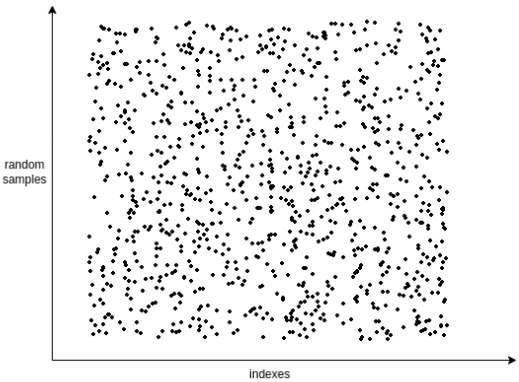}
	\caption{Scatter Graph for proposed PRNG}
	\label{fig:Scatter_Graph_for_proposed_PRNG}
\end{figure}

The scatter graph for the PRNG in the $Figure$ \ref{fig:Scatter_Graph_for_proposed_PRNG} is an embedding a functions that maps random numbers generated from this work to $3$ bit indexes $0-7$. The graph
clearly suggests that the random numbers generated are distributed uniformly along this space with no cluster formation.

\section{Comparison with other PRNGs}
This section presents a comprehensive comparison between the proposed PRNG and alternative approaches in terms of randomness, security, and speed.\\
The proposed PRNG is designed based on a combination of a lattice-based hard problem and LFSR, resulting in the generation of a random bit sequence with high effectiveness.\\
One notable advantage of the proposed scheme over other PRNG approaches is its ability to demonstrate desirable statistical properties and exhibit complete randomness. Furthermore, this study includes a comprehensive statistical analysis of the proposed PRNG scheme.\\
Table \ref{tab:prng-comparison} provides a detailed comparative analysis, comparing the suggested strategy with other similar approaches. The table evaluates various criteria and features across different PRNG methods, offering a thorough examination of the effectiveness of the proposed approach.\\
The proposed PRNG in this work was implemented on an Intel(R) Core(TM) i7-9700 processor. The system was equipped with 8 GB of RAM. The implementation was carried out in the C programming language, leveraging the computational capabilities and resources provided by the hardware setup.\\
Please note that the specific details of Table 1, including the criteria and comparative results, need to be filled in with appropriate information based on the actual comparison being made. 

\begin{table}[H] 
\small
\centering
\caption{Comparison of Proposed PRNG and Existing PRNGs}
\label{tab:prng-comparison}
\begin{tabular}{|m{1.9 cm} |m{1.5cm} |m{1.5cm} |m{1.3cm} |m{1.4cm} |m{1.3cm} |m{1.3cm}| }

 \toprule
\textbf{Features}                & \textbf{Proposed PRNG} & \textbf{\cite{b23-1}} & \textbf{\cite{b23-2}} & \textbf{\cite{b24}} & \textbf{\cite{quantis2014} QRNG} & \textbf{\cite{hotbits} QRNG} \\ 
\midrule
1) Implementation based on & Lattice-based hard problem & Linear congruential generator & Mersenne Twister & XORShift &  Quantum measurement of photons &  Utilizes radioactive decay\\
\hline
2) Randomness test   &  Die-Harder &  Die-Harder & - &  Die-Harder & Die-Harder & Die-Harder  \\
\hline
3) Key Space Analysis  & \checkmark  & \checkmark  &  \checkmark & \checkmark & - & - \\
\hline
4) Speed (Mbit/second) & 33.109 & 30.26 & 29.73  & 10.87 & 16 & 4 \\
\hline
5) Quantum Safe  & \checkmark & Not secure & Not secure  & Not secure & \checkmark & \checkmark \\
 \bottomrule
\end{tabular}
\begin{tablenotes}
		\footnotesize
		\item [1] \checkmark denotes "achieved," and "-" means there is no reported result
        \end{tablenotes}
\end{table}
A key feature that sets the proposed PRNG apart is its utilization of a post-quantum secure primitive, ensuring resilience against quantum attacks. The Linear Congruential Generator (LCG) \cite{b23-1} is known to have some weaknesses in terms of its statistical properties and predictability. It is vulnerable to certain types of attacks, such as the ‘state recovery attack’ and ‘period-finding attack’, which can compromise the security and unpredictability of the generated pseudorandom sequence. The Mersenne Twister \cite{b23-2} is a widely used pseudorandom number generator known for its long period. However, it is not designed to withstand attacks from quantum computers. The XORShift algorithm \cite{b24}, is not considered quantum safe. Quantum computers have the potential to break many classical PRNGs, including those based on XORShift. Considerable pre-processing is essential to prepare the random numbers generated by the QRNG \cite{quantis2014,hotbits} for usability. However, this pre-processing stage creates a bottleneck that limits the rate at which random numbers can be generated.\\
Please note that the reported speeds may vary depending on factors such as hardware, software implementation, and specific optimizations. The speeds provided here are approximate values for reference purposes.

\section{Application of Proposed PRNG in QKD}
QKD works by transmitting millions of polarized light particles (photons) over a ﬁber optic  cable  from  one  entity  to  another.  Polarization  is  measured  in  any  basis: two directions at right angles like rectilinear: horizontal and vertical. if a photon is polarized in a given basis it should be measured by the same basis else the measurement result is random. Each photon has a random quantum state, and collectively all the photons create a bit stream of ones and zeros. When the photons arrive at the endpoint, the receiver uses beam splitters (horizontal/vertical and diagonal) to "read" the polarization of each photon. The receiver does not know which beam splitter to use for each photon and  has  to  guess  which  one  to  use.  After  the  receiver  tells  the  sender  which  beam splitter was used for each of the photons in the sequence they were sent, the sender then compares that information with the sequence of polarizers used to send the photons. The photons that were read using the wrong beam splitter are discarded, and the resulting sequence of bits becomes a unique optical key that can be used to encrypt data. The assignment of basis from Alice is currently done by the use of QRNG and similarly Bob uses his QRNG to guess the basis sent by Alice. The obvious bottlenecks of QRNG slows down the process of QKD by slow generation of random bits hence slow assignment and guessing on both sides. However, the security of Alice and Bob is achieved only when both sides generate non deterministic random sequences of 0s and 1s.\\

\begin{figure}[H]
		\centering
		\includegraphics[height=12cm, width=11.5cm]{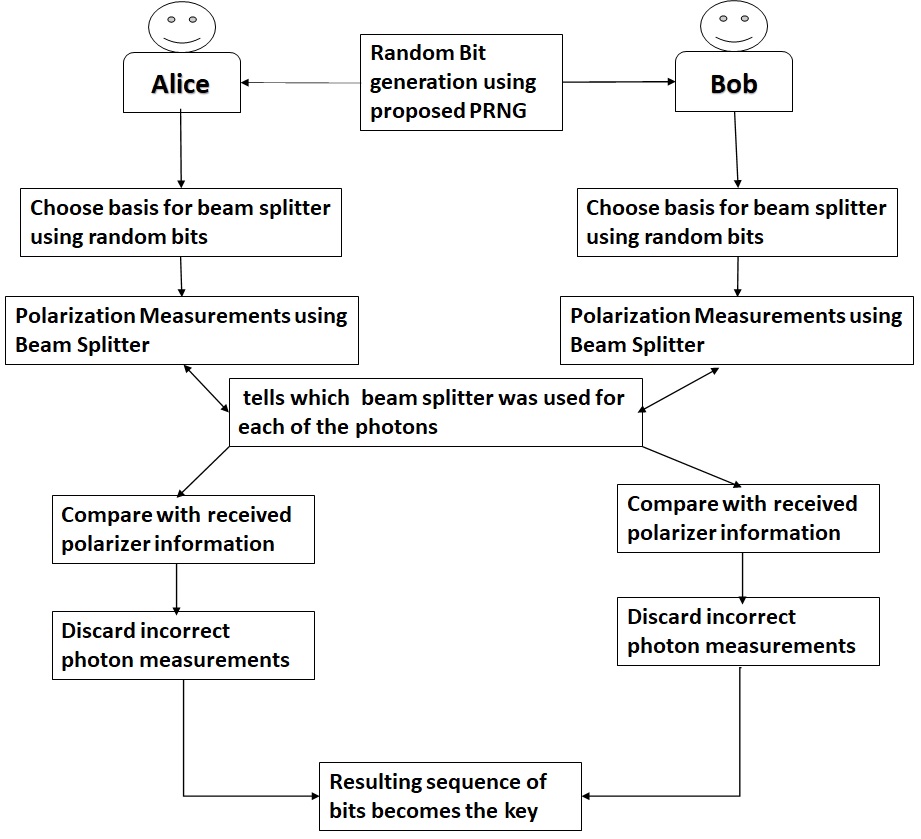}
		\caption{Key distribution in the quantum channel using the random bit generated from proposed PRNG}
		\label{fig:qkd_PRNG}
	\end{figure}

Our work achieves nondeterminism through the use of LWE function in our seed. The  Gaussian  error  distribution  in  LWE  problem  contributes  to  nondeterminism. Hence, as depicted in Fig. \ref{fig:qkd_PRNG}, the proposed PRNG generates completely different random sequences of bits for both Alice and Bob. Alice utilizes the proposed PRNG to generate random numbers and assigns her bits a random basis based on the PRNG sequence. On the other hand, Bob generates random sequences from the same PRNG to make guesses about the basis assigned by Alice.

\section{Conclusion}
The obvious bottlenecks of QRNG slow down the QKD process by producing random bits slowly, which leads to slow assignment and guessing on both sides, sender, and receiver. Additionally, the QRNG-generated sequence had a non-uniform distribution. Therefore, a different strategy was needed to address the inadequacies. 
The proposed method secured the seed by using lattice-based primitives, particularly the LWE problem. It generates an endless long random sequence using LFSR, satisfying the theoretical security criterion of the LWE problem. The generated sequence is uniformly distributed with improved speed. Die-Harder testing is applied to the proposed PRNG to test the randomness of the generated sequence and all 12 tests were passed.\\
 Non-deterministic PRNGs play a crucial role in QKD protocols to generate random keys. The future scope of PRNGs for QKD involves advancements in both theoretical and practical aspects. This includes developing more efficient and more secure algorithms for generating random keys, investigating techniques to enhance the entropy source, analyzing the impact of hardware limitations, and designing protocols that can withstand potential attacks in practical scenarios. Additionally, integrating PRNGs with emerging quantum technologies, such as quantum repeaters and quantum memories, can further enhance the security and scalability of QKD systems.

\section*{Acknowledgment}
  \noindent This research was supported by Defence Institute of Advanced Technology, Pune, India. We thank our colleagues who provided insight and expertise that greatly assisted the research of this paper.

\section*{Funding}
This research received grant from Defence Institute of Advanced Technology, Pune, India.

\bibliographystyle{plain}
\bibliography{cas-refs}
\end{document}